\documentstyle[prd,eqsecnum,aps,12pt]{revtex}

\input epsf

\begin{document}
\title{High temperature ${\rm ln}\left(T\right)$ contributions\\
in thermal field theory}

\author{F. T. Brandt and J. Frenkel}
\address{Instituto de F\'\i sica, Universidade de S\~ao Paulo,
S\~ao Paulo, 05389-970 SP, Brasil}

\date{\today}

\maketitle

\vskip 1.0cm

\begin{abstract}
We calculate, to one-loop order, the ${\rm ln}\left(T\right)$
contributions of 3-point functions in the $\phi^3$ and 
Yang-Mills theory at high temperature. We find that these
terms are Lorentz invariant and have the same structure as
the ultraviolet divergent contributions which occur at zero temperature.
A simple argument, valid for all $N$-point Green functions, is given for
this behavior.
\end{abstract}


\section{Introduction}

In thermal field theory one is often interested in the ``hard thermal''
loop contributions, meaning those terms which come from a  region where 
the loop momenta are of the order of the temperature $T$,
which is much larger than all external momenta. These Green functions,
which have a leading behavior at high temperature proportional to $T^2$,
where much studied in QCD 
\cite{Weldon,KajantieKapusta,Pisarski,BraatenPisarski92,FrenkelTaylor90}. 
They are an important tool
in {\it re-summing} the QCD thermal perturbation theory 
\cite{BraatenPisarski90}.
The hard thermal region is also relevant for the determination of the 
${\rm ln}\left(T\right)$ contributions (unlike the terms linear
in $T$ which come also from soft loop momenta). There have been several
investigations of the ${\rm ln}\left(T\right)$ terms associated with the
2-point functions 
\cite{Weldon,Kapusta,BrandtFrenkelTaylor91},
but the corresponding calculations done
in connection with the 3-point functions have been thus far restricted to
particular configurations of their external momenta
\cite{NakkagawaNiegawa,FujimotoYamada,BaierPireSchiff,EijckStephensWeert}.

This work intends to study the hard thermal ${\rm ln}\left(T\right)$
contributions associated with general
{\it 3-point Green functions} in thermal field theory.
In order to discuss the logarithmic dependence, we use the analytically 
continued imaginary-time perturbation theory \cite{Kapusta} and relate these
functions to {\it forward scattering amplitudes} 
\cite{FrenkelTaylor90,Barton}. 
In section II, we consider the 3-point function in the $\phi^3$ model in 
6 dimensions,
which has several similarities with the Yang-Mills theory and is useful
to illustrate the main points in a simpler context. In section III we
study the hard thermal behavior of the 3-point function in the
Yang-Mills theory. We focus on the ${\rm ln}(T)$ terms,
and find a Lorentz invariant result having a $T$ dependence 
which is directly connected with the ultraviolet structure of the
3-point function at {\it zero temperature}.
This conclusion, obtained in the Feynman gauge, is in fact true
in any gauge. 
(We have explicitly verified this statement in a general class of 
covariant gauges. Since this calculation requires a generalization
of the method of forward scattering amplitudes, the corresponding
analysis will be reported elsewhere.)
In general, the dependence upon $T$ for high $T$ is not necessarily related
to the ultraviolet divergence of the zero temperature amplitude. In the
Yang-Mills theory, for example, the leading behavior at high temperatures
for all $N$-gluon functions is proportional to $T^2$, although these functions
are ultraviolet finite for $N>4$. Hence, the above connection between the
${\rm ln}(T)$ contributions and the ultraviolet behavior at zero temperature,
which emerged after a rather involved calculation, 
may seem at first somewhat surprising.
For this reason we present
in section IV a general argument concerning the 
${\rm ln}\left(T\right)$ behavior of $N$-point gluon Green
functions. This gives a simple explanation for the fact that the 
${\rm ln}(T)$ contributions always appear with the same coefficient as the
residue of the ultraviolet pole part of the zero temperature amplitude.

\section{The 3-point scalar function}\label{sec2}

In order to exhibit the behavior of sub-leading hard thermal contributions
in the simplest way, we consider the massless
$\phi^3$ model in 6 dimensions which 
is asymptotically free. The Feynman diagram associated with the thermal
3-point function is shown in Fig. 1a. The analytically continued imaginary
time perturbation theory can be formulated 
\cite{FrenkelTaylor90,Barton} so as to express this 
function in terms of forward scattering amplitudes of on-shell particles,
as illustrated in Fig. 1b. Here $Q =(|{\bf Q}|,{\bf Q})$ is the 
four-momentum of the on-shell thermal particle. There are 6 diagrams
such as this one, which are obtained by permutations of the external
momenta $k_i$. These contributions can be written as
\begin{equation}\label{eq21}
\Gamma_3=
\frac{\lambda^3}{\left(2\pi\right)^5}\int\frac{{\rm d}^5Q}{2|{\bf Q}|}
\left\{\left[\frac{1}{2Q\cdot k_1+k_1^2}\;
\frac{1}{-2Q\cdot k_2+k_2^2}
+permutations\right]+Q\rightarrow -Q\right\}N(|{\bf Q}|),
\end{equation}
where $N(|{\bf Q}|)$ is the Bose-Einstein distribution
\begin{equation}\label{eq22}
N(|{\bf Q}|)=\frac{1}{\exp{\left(|{\bf Q}|/T\right)}-1}
\end{equation}
  \begin{figure}[htb]
   \hspace{.1\textwidth}

   \vbox{
    \epsfxsize=.8\textwidth
    \epsfbox{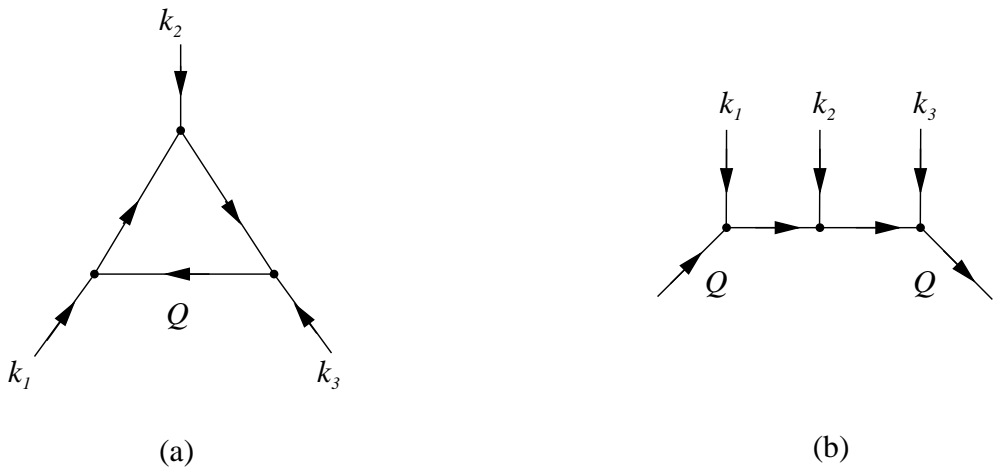}}

   \label{Fig1}
\caption{(a) The 3-scalar thermal loop diagram. Momentum and energy
conservation is understood at each vertex.
(b) An example of the forward scattering graph connected with diagram (1a).
}
  \end{figure}
Since in the hard thermal region we require large values of $|{\bf Q}|$,
we may expand each denominator as
\begin{equation}\label{eq23}
\frac{1}{2Q\cdot k_i+k_i^2}=\frac{1}{2Q\cdot k_i}- 
\frac{k_i^2}{\left(2Q\cdot k_i\right)^2}+
\frac{\left(k_i^2\right)^2}{\left(2Q\cdot k_i\right)^3}+\cdots .
\end{equation}
The first term has each denominator of the form 
$\left(Q\cdot k_i\right)^{-1}$.
Such terms would give individually $T^2$ contributions,but these actually
cancel by the eikonal identity
\begin{equation}\label{eq24}
\frac{1}{Q\cdot k_1 Q\cdot k_2}+
\frac{1}{Q\cdot k_1 Q\cdot k_3}+
\frac{1}{Q\cdot k_2 Q\cdot k_3}=0
\end{equation}
The second term in (\ref{eq23}) is down by one power of $k_i/|{\bf Q}|$ and
would lead to individual $T$ contributions, but such terms also cancel
by symmetry under $Q\rightarrow -Q$. Hence, we are left with the following
sub-leading contributions
\begin{eqnarray}\label{eq25}
\Gamma_3 & = &
\frac{\lambda^3}{16\left(2\pi\right)^5}\int \frac{{\rm d}^5 Q}{|{\bf Q}|}
N\left(|{\bf Q}|\right)\nonumber \\
{} & \times &
\left\{\frac{1}{Q\cdot k_2 Q\cdot k_3}\left[
\frac{k_1^2\,k_2^2}{Q\cdot k_1 Q\cdot k_2}-
\frac{\left(k_1^2\right)^2}{\left(Q\cdot k_1\right)^2}-
\frac{\left(k_2^2\right)^2}{\left(Q\cdot k_2\right)^2}\right]+
permutations\right\}.
\end{eqnarray}
After some algebra, which makes use of the eikonal identity (\ref{eq24}) and
the overall momentum conservation of the external momenta, the above expression
can be reduced to the form
\begin{equation}\label{eq26}
\Gamma_3=
\frac{\lambda^3}{8\left(2\pi\right)^5}\int \frac{{\rm d}^5 Q}{|{\bf Q}|}
N\left(|{\bf Q}|\right)
\left\{
\frac{k_1^2\,k_2\cdot k_3}{\left(Q\cdot k_1\right)^2 Q\cdot k_2 Q\cdot k_3}+
permutations\right\},
\end{equation}
which involves only homogeneous functions of zero degree in each of the
external momenta. In order to perform this integral, it is convenient to
rewrite Eq. (\ref{eq26}) as
\begin{equation}\label{eq27}
\Gamma_3=
\frac{\lambda^3}{8\left(2\pi\right)^5}\int^\infty 
\frac{{\rm d} |{\bf Q}|}{|{\bf Q}|}
N\left(|{\bf Q}|\right)\int {\rm d}\Omega\left\{
\frac{k_1^2\,k_2\cdot k_3}{\left(\hat Q\cdot k_1\right)^2 
\hat Q\cdot k_2 \hat Q\cdot k_3}+ permutations\right\},
\end{equation}
where 
$\int {\rm d}\Omega$ denotes angular integration of the 5-dimensional
unit vector $\hat {{\bf Q}}$ and $\hat Q \equiv(1,\hat {{\bf Q}})$.

We have evaluated the angular integral, using the approach discussed in
reference \cite{FrenkelTaylor90}. 
From the fact that the integrand is a
dimensionless function of zero degree in the external momenta, one
finds after some calculation that the angular integration just gives a
factor of $8\pi^2$. 
In the integration over ${\rm d} |{\bf Q}|$, the lower limit
must be chosen consistently with the range of validity of the expansion
in Eq. (\ref{eq23}). However, its value is immaterial for the determination
of the ${\rm ln}\left(T\right)$ 
contribution, which comes from the region of high internal momenta
$|{\bf Q}|\sim T\gg k_i$. Evaluating
these terms, we then obtain for the logarithmic dependence on the
temperature the simple Lorentz invariant result
\begin{equation}\label{eq28}
\tilde\Gamma_3=-\frac{\lambda^3}{64 \pi^3}{\rm ln}\left(T\right) .
\end{equation}
This equation may be compared with the ultraviolet divergent contribution
of the 3-point function at zero temperature, evaluated in
$6-2\epsilon$ dimensions:
\begin{equation}\label{eq29}
\Gamma_3^{UV}=\frac{\lambda^3}{64 \pi^3}\frac{M^{2\epsilon}}{2\epsilon}
\simeq
\frac{1}{64 \pi^3}\left[\frac{1}{2\epsilon}+{\rm ln}\left(M\right)\right],
\end{equation}
where  $M$ is the renormalization scale.
We see that the ${\rm ln}\left(T\right)$ contribution has the same
structure as the ultraviolet divergent
part at zero temperature, so that it
may be naturally combined with the ${\rm ln}\left(M\right)$ term.
The same behavior was previously noted in connection with the scalar
self-energy function \cite{BrandtFrenkelTaylor91}.

\section{The 3-gluon function}\label{sec3}
It is known that the above connection between the ${\rm ln}(T)$
contributions at high temperature and the ultraviolet
pole part at zero temperature is also exhibited by the two point gluon
function in the Yang-Mills theory \cite{Weldon,BrandtFrenkelTaylor91}. 
If this property continues to
hold for the 3-point gluon function, then,  using
the well known result for the corresponding zero temperature
amplitude, we could immediately write for the ${\rm ln}(T)$ 
contributions the following {\it ansatz}:
\begin{equation}\label{eq31a}
\tilde\Gamma_{3\;\mu_1\mu_2\mu_3}^{a_1 a_2 a_3}=
\frac{1}{12\pi^2}g^2\,V_{\mu_1\mu_2\mu_3}^{a_1 a_2 a_3}\;
{\rm ln}\left(T\right),
\end{equation}
where $a_1,\, a_2,\,a_3$ are color indices, $\mu_1,\, \mu_2,\, \mu_3$
are Lorentz indices and 
\begin{equation}\label{eq31b}
V_{\mu_1\mu_2\mu_3}^{a_1 a_2 a_3}=
-i g f^{a_1 a_2 a_3}\left[\eta_{\mu_1 \mu_2} (k_1-k_2)_{\mu_3} +
                          \eta_{\mu_2 \mu_3} (k_2-k_3)_{\mu_1} +
                          \eta_{\mu_3 \mu_1} (k_3-k_1)_{\mu_2}\right] 
\end{equation}
is the basic three gluon vertex. 
Equation (\ref{eq31a}) was obtained
replacing $1/(2\epsilon)$ by ${\rm ln}(1/T)$ in the $T=0$
ultraviolet divergent 3-gluon function, computed in the {\it Feynman gauge}.
In what follows we will prove that this {\it ansatz} is indeed correct.

In the Feynman gauge the {\it pole structure} 
of the gluon propagator is very similar to the scalar case considered in 
the previous section; there are only {\it simple poles}. 
This is a important ingredient 
which makes it possible to formulate the imaginary time formalism in 
terms of {\it forward scattering amplitudes} 
\cite{FrenkelTaylor90,Barton}.
It is then straightforward to write the  following expression for the 
3-gluon function
\begin{equation}\label{eq31}
\Gamma_{\mu_1\mu_2\mu_3}^{a_1 a_2 a_3}\left(k_1,k_2,k_3\right)=
\frac{g^3}{\left(2\pi\right)^3}\int\frac{{\rm d}^3 {\bf Q}}
{2|{\bf Q}|} N\left(|{\bf Q}|\right)
\left[{\cal S}_{\mu_1\mu_2\mu_3}^{a_1 a_2 a_3}
\left(k_1,k_2,k_3,Q\right) + Q\rightarrow -Q \right],
\end{equation}
where ${\cal S}_{\mu_1\mu_2\mu_3}^{a_1 a_2 a_3}\left(k_1,k_2,k_3,Q\right)$
is the forward scattering amplitude given by the sum of the diagrams shown
in Figs. (2b) and (2d) and the graphs obtained by 
permutations of their external momenta and indices.
We can now use a hard thermal expansion like Eq. (\ref{eq23}) for 
the denominators in 
${\cal S}_{\mu_1\mu_2\mu_3}^{a_1 a_2 a_3}\left(k_1,k_2,k_3,Q\right)$.
  \begin{figure}[htb]
   \hspace{.1\textwidth}

   \vbox{
    \epsfxsize=.8\textwidth
    \epsfbox{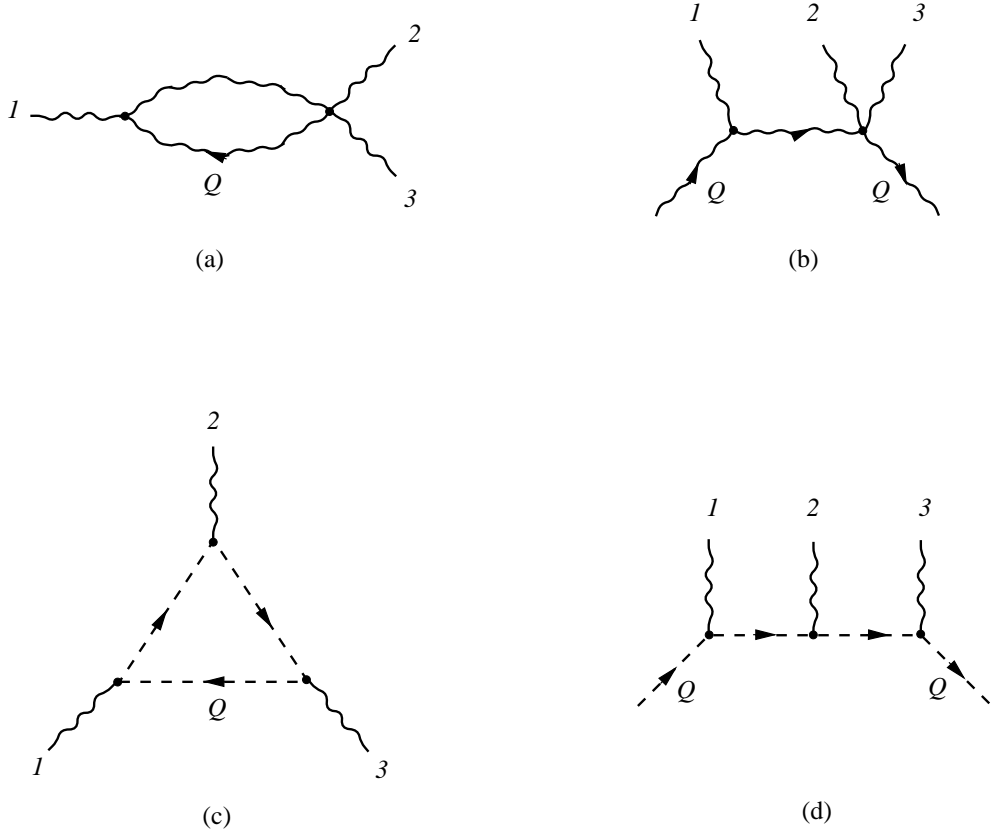}}
   \label{Fig2}
\caption{
The thermal 3-gluon loop diagram.
Wavy lines represent {\it gluons}
and the broken lines in (c) and (d) denote {\it either} ghosts 
{\it or} internal gluons.
The numbers 1, 2, and 3
represent a collective index for the momenta, Lorentz and color indices.
Graphs (b) and (d) are examples of 
forward scattering amplitudes associated, respectively  with
the diagrams (a) and (c).
All external momenta are inward and $k_1+k_2+k_3=0$.}
  \end{figure}
As in the scalar case,
odd powers of $Q$ will cancel when the $Q\rightarrow -Q$ terms are
added. From the momentum dependence of the Yang-Mills vertices it is easy
to see that the resulting terms in the hard thermal expansion of
${\cal S}_{\mu_1\mu_2\mu_3}^{a_1 a_2 a_3}\left(k_1,k_2,k_3,Q\right)$
will be functions of decreasing degree in $Q$ starting from zero
degree. 
Therefore, by naive power counting in Eq. (\ref{eq31})
we conclude that the terms of degree $0$ and $-2$ in
${\cal S}_{\mu_1\mu_2\mu_3}^{a_1 a_2 a_3}\left(k_1,k_2,k_3,Q\right)$
will produce respectively
the leading $T^2$ and the ${\rm ln}(T)$ contributions.
The leading $T^2$ contribution does not vanish as in the
$\phi^3$ model of the previous section and the result is well
known \cite{BraatenPisarski92,FrenkelTaylor90,BraatenPisarski90}. 
Using a similar procedure as the one in
(\ref{eq27}), we get for the ${\rm ln}(T)$ contributions an expression
of the form
\begin{equation}\label{eq32}
\tilde\Gamma_{3\;\mu_1\mu_2\mu_3}^{a_1 a_2 a_3}\left(k_1,k_2,k_3\right) =
-\displaystyle{ \frac{g^3}{4\,\pi^2}}
\displaystyle{{\rm ln}\left({T}\right)}
\displaystyle{\int \frac{{\rm d}\Omega}{4\pi}}\;\;
{\cal L}_{\mu_1\mu_2\mu_3}^{a_1 a_2 a_3}
\left(k_1,k_2,k_3,\hat Q\right)\nonumber ,
\end{equation}
where $\int {\rm d} \Omega$ denotes angular integration of the
3-dimensional unit vector 
$\hat {{\bf Q}}$, $\hat Q \equiv(1,\hat {{\bf Q}})$
and ${\cal L}_{\mu_1\mu_2\mu_3}^{a_1 a_2 a_3}
\left(k_1,k_2,k_3,\hat Q\right)$ is a function of degree $-2$ in $\hat Q$.

All angular integrals of the form 
$\int {\rm d} \Omega {\cal L}_{\mu_1\mu_2\mu_3}^{a_1 a_2 a_3}$
can be generated differentiating the following basic integral
with respect to ${k_i}_{\mu_l}$
\cite{FrenkelTaylor90}
\begin{equation}\label{eq33}
\begin{array}{lll}
\displaystyle{\int}
\displaystyle{\frac{{\rm d} \Omega}{4\,\pi}} 
\displaystyle{\frac{1}{k_i\cdot \hat Q\; k_j\cdot\hat Q}}
& = &
\displaystyle{\int_0^1}{\rm d} x 
\displaystyle{\int }\displaystyle{\frac{{\rm d} \Omega}{4\,\pi}}
\left\{\left[x k_i+(1-x) k_j\right]\cdot \hat Q\right\}^{-2} \\
{} & {} & {} \\
& = & 
\displaystyle{\int_0^1}{\rm d} x
\left[x k_i+(1-x) k_j\right]^{-2}.
\end{array}
\end{equation}
The right hand side of (\ref{eq33}) was obtained using standard
{\it Feynman parameterization}. After an elementary angular 
integration, a manifest {\it Lorentz scalar} is unveiled in the resulting 
Feynman parameter integral which can be explicitly performed
\cite{FrenkelTaylor90}.
Thus, the {\it Lorentz covariance} of the logarithmic contributions
is established. 
This remarkable property, which holds in spite of the fact that the
angular integral is not a generally Lorentz invariant process,
was shown to be true in reference 
\cite{FrenkelTaylor90} for {\it any} 
angular integrand which is a function of degree $-2$ in $\hat Q$.
We also note that, by power counting, the ${\rm ln}(T)$ contributions
of {\it any} hard thermal $N$-point function will equally be
a Lorentz invariant quantity.
In what follows we will use this property as an important tool
for the explicit computation of the ${\rm ln}(T)$ contributions.

Our simple {\it ansatz} given by Eq. (\ref{eq31a}) is a very special
case of the most general Lorentz covariant expression.
In general, for each set of color indices,
there are 14 tensor structures which can be formed using 3 Lorentz indices and 
the 2 independent four-momenta. However, the explicit calculation of the diagrams in Fig. 2 shows 
that the angular integrals in (\ref{eq32}) are such that the totally 
antisymmetric color factor $f^{a_1\,a_2\,a_3}$ {\it factorizes} and the result
can be written as
\begin{equation}\label{eq34}
\int \frac{{\rm d}\Omega}{4\,\pi}\;\;
{\cal L}_{\mu_1\mu_2\mu_3}^{a_1 a_2 a_3}\left(k_1,k_2,k_3,\hat Q\right)=
-i\;f^{a_1\,a_2\,a_3}\;{\cal A}_{\mu_1\mu_2\mu_3}
\left(k_1,k_2,k_3\right).
\end{equation}
Therefore, the Lorentz covariant tensor 
${\cal A}_{\mu_1\mu_2\mu_3}\left(k_1,k_2,k_3\right)$
must be {\it antisymmetric} under any interchange of a pair of momenta and the 
corresponding Lorentz indices. A straightforward but quite involved
calculation (we have used the {\it symbolic computer program} Maple) 
yields
\begin{equation}\label{eq35}
{\cal A}_{\mu_1\mu_2\mu_3}\left(k_1,k_2,k_3\right)\equiv
\int \frac{{\rm d}\Omega}{4\,\pi}\;
{\cal L}_{\mu_1\mu_2\mu_3}\left(k_1,k_2,k_3,\hat Q\right),
\end{equation}
where the tensors
${\cal L}_{\mu_1\mu_2\mu_3}\left(k_1,k_2,k_3,\hat Q\right)$
are given in the appendix. 

We can perform the angular integrals 
for each individual term of these expressions 
in a straightforward way,
using for instance the procedure of differentiation with respect to 
${k_i}_{\mu_l}$ as described above. In practice, we found easier 
to use a decomposition in terms 
of a set of tensors \cite{DavydychevOsland,BallChiu} built from
the basic tensors

\begin{equation}
\begin{array}{lll}
A^1_{\mu_1\mu_2\mu_3}\left(k_1,\,k_2,\,k_3\right)&=&
\eta_{\mu_1 \mu_2} (k_1-k_2)_{\mu_3} 
\\ 

A^2_{\mu_1\mu_2\mu_3}\left(k_1,\,k_2,\,k_3\right)&=&
\left({k_1\cdot k_2}\;\eta_{\mu_1 \mu_2} - {k_1}_{\mu_2} {k_2}_{\mu_1}\right)
\; (k_1-k_2)_{\mu_3}
\\ 

A^3_{\mu_1\mu_2\mu_3}\left(k_1,\,k_2,\,k_3\right)&=&
\left(k_1\cdot k_2\;\eta_{\mu_1 \mu_2} - {{k_1}_{\mu_2} {k_2}_{\mu_1}} \right)
\; \left( {k_1}_{\mu_3}{k_2\cdot k_3} - 
          {k_2}_{\mu_3}{k_1\cdot k_3} \right)
\\ 

A^4_{\mu_1\mu_2\mu_3}\left(k_1,\,k_2,\,k_3\right)&=&
\eta_{\mu_1 \mu_2} 
   \left( {k_1}_{\mu_3}{k_2\cdot k_3} - 
          {k_2}_{\mu_3}{k_1\cdot k_3} \right)
\\
{}&{}&+\frac 1 3 {\left( {k_1}_{\mu_3} {k_2}_{\mu_1} {k_3}_{\mu_2}
        - {k_1}_{\mu_2} {k_2}_{\mu_3} {k_3}_{\mu_1} \right)}
\\ 

S^1_{\mu_1\mu_2\mu_3}\left(k_1,\,k_2,\,k_3\right)&=&
\eta_{\mu_1 \mu_2} (k_1+k_2)_{\mu_3} 
\\ 

S^2_{\mu_1\mu_2\mu_3}\left(k_1,\,k_2,\,k_3\right)&=&
 {k_1}_{\mu_3} {k_2}_{\mu_1} {k_3}_{\mu_2}
+{k_1}_{\mu_2} {k_2}_{\mu_3} {k_3}_{\mu_1}.
\end{array}
\end{equation}
The {\it complete} set is generated from the above equations including
new tensors obtained from
$A^i_{\mu_1\mu_2\mu_3}\left(k_1,\,k_2,\,k_3\right)$, $i=1,\,2,\,3,\,4$, 
and from 
$S^j_{\mu_1\mu_2\mu_3}\left(k_1,\,k_2,\,k_3\right)$, $j=1,\,2$,
by cyclic permutations
of $(k_1,\mu_1),\, (k_2,\mu_2),\, (k_3,\mu_3)$. This gives a total of 16
tensors in terms of which we write the most general expression for 
${\cal A}_{\mu_1\mu_2\mu_3}\left(k_1,k_2,k_3\right)$. 
The coefficients of this expansion can be obtained in a straightforward
way by simply solving a system of 16 equations. These equations are
obtained using the expressions in the Appendix and performing the Lorentz
indices contractions with each of the 16 tensors.
The solution of this system will be, in general, scalar functions of 
$k_1,\,k_2,\,k_3$ involving angular integrals which can always be
reduced to Eq. (\ref{eq33}), or the special case of it when $k_i=k_j$,
when one uses the momentum conservation $k_1+k_2+k_3=0$.
In principle these scalars could have any kind of dependence
on the external momenta involving Eq. (\ref{eq33})
and rational functions. However, the explicit calculation shows that, after 
using the eikonal identity given by Eq. (\ref{eq24}), all 
coefficients vanish, except for the coefficients of 
$A^1_{\mu_1\mu_2\mu_3}\left(k_1,\,k_2,\,k_3\right)=
\eta_{\mu_1 \mu_2} (k_1-k_2)_{\mu_3}$ and its cyclic permutations
which simplify to give just $-1/3$. 
Inserting these terms into Eq (\ref{eq34}) and using Eq. (\ref{eq32}),
we finally obtain the result stated in
Eq. (\ref{eq31a}). 

\section{Discussion}\label{discussion}

To get a further understanding of the connection between  the 
${\rm ln}(T)$ contributions and the ultraviolet
behavior of the Green functions at zero temperature, let us
consider the complete thermal amplitude, which includes the zero temperature
part. This can be written, for instance in the Yang-Mills case 
where we omit for simplicity the color indices,
as follows \cite{Kapusta,BrandtFrenkel}:
\begin{equation}
A_{\mu_1\cdots\mu_N}(k_{i0},{\bf k}_i,T)=
M^{2\epsilon} \,T \sum_{Q_0=2\pi n i T} \int {\rm d}^{3-2\epsilon}
{\bf Q} F_{\mu_1\cdots\mu_N}(Q_0,{\bf Q},k_{i0},{\bf k}_i).
\end{equation}
Here $M$ is the renormalization scale, $k_{i0}/2\pi i T$ are 
integers
and $n$ runs over all integers. For fixed $n$, the ${\bf Q}$ integral is
ultraviolet finite, having no poles at $\epsilon=0$.

In order to determine those terms which can yield an ultraviolet pole when the
summation over $n$ is evaluated, let us examine a relevant contribution which
is obtained after the ${\bf Q}$ integration has been performed. 
Making appropriate shifts in $Q_0$, one finds that such a term is proportional
to a sum of the form
\begin{equation}\label{eq42}
S=\sum_{Q_0=2\pi n i T}
\frac{1}{\left(Q_0^2+ a k_0^2 + b {\bf k}^2\right)^{\frac 1 2
+\epsilon}},
\end{equation}
where $a$, $b$ are  constants and $k_0$ is some linear combination of the 
external energies with integral coefficients. ${\bf k}$ is some linear
combination of the external momenta which may be neglected in the high
temperature limit, except when $Q_0^2+a k_0^2$ vanishes.

We now set $k_0=2\pi l i T$, 
where $l$ is some integer and consider the contributions
to $S$ from the regions 
$n^2< l^2$ and $n^2> l^2$. It turns out that the pole part
arises from the summation over the domain where $|n|\gg|l|$, i. e., when
$|Q_0|\gg|k_0|$. Expanding in this region (\ref{eq42}) in powers of 
$Q_0^{-1}$, we find that the leading term gives a contribution involving
the {\it zeta function} $\zeta(1+2\epsilon)$, which is defined in
general as \cite{Gradshteyn}
\begin{equation}
\zeta(z)=\sum_{n=1}^{\infty}\frac{1}{n^z}.
\end{equation}
This function is analytic for all values of $z$, except near the point
$z=1+2\epsilon$, where it has a simple pole $1/2\epsilon$.

At this stage, having performed the summation over the discrete frequencies 
$Q_0=2\pi n i T$, we can analytically continue the external energies
to continuous values of $k_0$.
Identifying in the complete thermal amplitude all contributions
which yield poles at $\epsilon=0$, 
and using the fact that the leading term
in $S$ is proportional to $T^{-1-2\epsilon}$,
one obtains an expression of the form:
\begin{eqnarray}
A^\epsilon_{\mu_1\cdots\mu_N}(k_i,T) & = &
\left(\frac{M}{T}\right)^{2\epsilon}\frac{1}{2\epsilon}
{\cal R}_{\mu_1\cdots\mu_N}(k_i) \nonumber \\ 
{} & \simeq {} &
\left(\frac{1}{2\epsilon}-{\rm ln}\frac T M\right)
{\cal R}_{\mu_1\cdots\mu_N}(k_i).
\end{eqnarray}
where ${\cal R}_{\mu_1\cdots\mu_N}$ is the residue of the ultraviolet
divergent part of the Green function at zero temperature. 
This has the same structure, because of the renormalizability of the
the theory, as the corresponding basic function appearing in the
Yang-Mills Lagrangian.

The above equation shows that for general Green functions, the 
${\rm ln}(T)$ contributions have the same form as
the ultraviolet divergent terms which occur at zero temperature
and combine in a simple way with the ${\rm ln}(M)$ terms.
In particular, if the Green function at zero temperature is ultraviolet
convergent, the residue ${\cal R}_{\mu_1\cdots\mu_N}$ must vanish and
the ${\rm ln}(T)$ term should be absent at high temperature.
This has been explicitly verified in the case of the electron-positron
box diagram in thermal QED \cite{BrandtFrenkelTaylor94}.

In conclusion, we mention that the above property allows us to include
in a simple way the ${\rm ln}\left(T\right)$ contributions into the
{\it running coupling constant} $g\left(T\right)$ at high temperature.
Several investigations on this important topic 
\cite{PerryCollins,MatsumotoNakanoUmezawa,ElmforsKobes,ChaichianHayashi}
have exposed ambiguities which are related, at least in part, to the fact
that the thermal contributions to $g\left(T\right)$ are not generally
Lorentz invariant functions. On the other hand, the ${\rm ln}\left(T\right)$
terms are Lorentz invariant, being directly related to the
ultraviolet behavior of the Green functions at zero temperature.
It is well known that the effective coupling $\bar g\left(\kappa/M\right)$
at zero temperature, where $\kappa$ is a typical external momentum,
involves a logarithmic dependence of the form 
${\rm ln}\left(\kappa/M\right)$. The $\kappa$ dependence in this term must be
canceled by a corresponding dependence in the 
${\rm ln}\left(T/\kappa\right)$ term at high temperature, 
since the complete thermal amplitude contains only a combination of the form
${\rm ln}\left(T/M\right)$. We thus obtain a gauge and Lorentz
invariant quantity $\bar g\left(T/M\right)$ which is relevant, for
example, in the calculation of the pressure in thermal field theories
\cite{FrenkelSaaTaylor,ArnoldZhai,BraatenNieto}.

\acknowledgements{We would like to thank CNPq (Brasil) for a grant. J. F. is 
grateful to Prof. J. C. Taylor for helpful discussions.}

\newpage
\appendix
\section{}\label{app1}

In this appendix we present the results for the integrand
of the angular integrals corresponding to the diagrams of Fig. 2.
In terms of the individual contributions of each diagram, we can
write

\begin{eqnarray}
{\cal L}_{\mu_1\mu_2\mu_3}\left(k_1,k_2,k_3,\hat Q\right) & = &
{\cal L}^{tadpole}_{\mu_1\mu_2\mu_3}\left(k_1,k_2,k_3,\hat Q\right)+
{\cal L}^{ghost}_{\mu_1\mu_2\mu_3}\left(k_1,k_2,k_3,\hat Q\right) 
\nonumber \\
{} & + &  
{\cal L}^{gluon}_{\mu_1\mu_2\mu_3}\left(k_1,k_2,k_3,\hat Q\right),
\end{eqnarray}
where

\begin{equation}
{\cal L}^{tadpole}_{\mu_1\mu_2\mu_3}\left(k_1,k_2,k_3,\hat Q\right)=
-{\frac {9\,{k_3\cdot k_3}\,k_{{3\mu_1}}\eta_{{\mu_2\mu_3}}}{8\,{ \left(\hat Q\cdot k_3\right)}^{2}}
}+{\frac {9\,{k_3\cdot k_3}\eta_{{\mu_1\mu_3}}k_{{3\mu_2}}}{8\,{ \left(\hat Q\cdot k_3\right)}^{2}
}}
\end{equation}
$$
+ \left\{ \; \mbox{cyclic permutations of} \; (k_1,\mu_1),
(k_2,\mu_2), (k_3,\mu_3)\; \right\}
$$

\begin{equation}
{\cal L}^{ghost}_{\mu_1\mu_2\mu_3}\left(k_1,k_2,k_3,\hat Q\right)=
\end{equation}
$$
-{\frac {k_{{1\mu_3}}k_{{2\mu_2}}\hat Q_{{\mu_1}}{k_2\cdot k_2}}{16\,{ \left(\hat Q\cdot k_2\right)}^{2}
 \left(\hat Q\cdot k_1\right)}}-{\frac {k_{{2\mu_2}}\hat Q_{{\mu_1}}\hat Q_{{\mu_3}}{\left(k_2\cdot k_2\right)}^{2}}{32\,{
 \left(\hat Q\cdot k_2\right)}^{3} \left(\hat Q\cdot k_1\right)}}-{\frac {\hat Q_{{\mu_2}}\hat Q_{{\mu_3}}\hat Q_{{\mu_{{1
}}}}{\left(k_2\cdot k_2\right)}^{2}{k_1\cdot k_1}}{32\,{ \left(\hat Q\cdot k_2\right)}^{3}{ \left(\hat Q\cdot k_1\right)}^{2}}
}
$$
$$
+{\frac {\hat Q_{{\mu_2}}\hat Q_{{\mu_3}}\hat Q_{{\mu_1}}{k_2\cdot k_2}\,{\left(k_1\cdot k_1\right)}^{2}}{32\,{ \left(\hat Q\cdot k_2\right)}^{2}{ \left(\hat Q\cdot k_1\right)}^{3}}}-{\frac {k_{{2\mu_3}}k_{{1\mu_1}}\hat Q_{{\mu_2
}}{k_2\cdot k_2}}{16\,{ \left(\hat Q\cdot k_2\right)}^{2} \left(\hat Q\cdot k_1\right)}}+{\frac {k_{{2\mu_3}}k
_{{1\mu_1}}\hat Q_{{\mu_2}}{k_1\cdot k_1}}{16\, \left(\hat Q\cdot k_2\right){ \left(\hat Q\cdot k_1\right)}^{2}
}}
$$
$$
-{\frac {k_{{2\mu_3}}\hat Q_{{\mu_2}}\hat Q_{{\mu_1}}{\left(k_1\cdot k_1\right)}^{2}}{32\, \left(\hat Q\cdot k_2\right){
 \left(\hat Q\cdot k_1\right)}^{3}}}+{\frac {k_{{2\mu_2}}\hat Q_{{\mu_1}}\hat Q_{{\mu_3}}{k_2\cdot k_2}\,
{k_1\cdot k_1}}{32\,{ \left(\hat Q\cdot k_2\right)}^{2}{ \left(\hat Q\cdot k_1\right)}^{2}}}-{\frac {k_{{2\mu_2}}\hat Q_
{{\mu_1}}\hat Q_{{\mu_3}}{\left(k_1\cdot k_1\right)}^{2}}{32\, \left(\hat Q\cdot k_2\right){ \left(\hat Q\cdot k_1\right)}^
{3}}}
$$
$$
-{\frac {k_{{2\mu_3}}\hat Q_{{\mu_2}}\hat Q_{{\mu_1}}{\left(k_2\cdot k_2\right)}^{2}}{32\,{ \left(\hat Q\cdot k_2\right)}^
{3} \left(\hat Q\cdot k_1\right)}}+{\frac {k_{{2\mu_3}}\hat Q_{{\mu_2}}\hat Q_{{\mu_1}}{k_2\cdot k_2}\,
{k_1\cdot k_1}}{32\,{ \left(\hat Q\cdot k_2\right)}^{2}{ \left(\hat Q\cdot k_1\right)}^{2}}}-{\frac {\hat Q_{{\mu_2}}\hat Q_{{
\mu_3}}\hat Q_{{\mu_1}}{\left(k_1\cdot k_1\right)}^{3}}{32\, \left(\hat Q\cdot k_2\right){ \left(\hat Q\cdot k_1\right)}^{4
}}}
$$
$$
+{\frac {k_{{1\mu_1}}\hat Q_{{\mu_2}}\hat Q_{{\mu_3}}{\left(k_2\cdot k_2\right)}^{2}}{32\,{ \left(\hat Q\cdot k_2\right)}^{
3} \left(\hat Q\cdot k_1\right)}}+{\frac {k_{{1\mu_1}}\hat Q_{{\mu_2}}\hat Q_{{\mu_3}}{\left(k_1\cdot k_1\right)}^{2}}{32
\, \left(\hat Q\cdot k_2\right){ \left(\hat Q\cdot k_1\right)}^{3}}}-{\frac {k_{{1\mu_1}}\hat Q_{{\mu_2}}\hat Q_{{
\mu_3}}{k_2\cdot k_2}\,{k_1\cdot k_1}}{32\,{ \left(\hat Q\cdot k_2\right)}^{2}{ \left(\hat Q\cdot k_1\right)}^{
2}}}
$$
$$
+{\frac {k_{{1\mu_3}}\hat Q_{{\mu_2}}\hat Q_{{\mu_1}}{\left(k_2\cdot k_2\right)}^{2}}{32\,{ \left(\hat Q\cdot k_2\right)}^{
3} \left(\hat Q\cdot k_1\right)}}-{\frac {k_{{1\mu_3}}\hat Q_{{\mu_2}}\hat Q_{{\mu_1}}{k_2\cdot k_2}\,
{k_1\cdot k_1}}{32\,{ \left(\hat Q\cdot k_2\right)}^{2}{ \left(\hat Q\cdot k_1\right)}^{2}}}+{\frac {k_{{1\mu_3}}\hat Q_
{{\mu_2}}\hat Q_{{\mu_1}}{\left(k_1\cdot k_1\right)}^{2}}{32\, \left(\hat Q\cdot k_2\right){ \left(\hat Q\cdot k_1\right)}^
{3}}}
$$
$$
+{\frac {\hat Q_{{\mu_2}}\hat Q_{{\mu_3}}\hat Q_{{\mu_1}}{\left(k_2\cdot k_2\right)}^{3}}{32\,{ \left(\hat Q\cdot k_2\right)}^{4}
 \left(\hat Q\cdot k_1\right)}}+{\frac {k_{{1\mu_3}}k_{{2\mu_2}}\hat Q_{{\mu_1}}{k_1\cdot k_1}}{16\,
 \left(\hat Q\cdot k_2\right){ \left(\hat Q\cdot k_1\right)}^{2}}}
$$
$$
+ \left\{ \; \mbox{cyclic permutations of} \; (k_1,\mu_1),
(k_2,\mu_2), (k_3,\mu_3)\; \right\}
$$

\begin{equation}
{\cal L}^{gluon}_{\mu_1\mu_2\mu_3}\left(k_1,k_2,k_3,\hat Q\right)=
\end{equation}
$$
+{\frac {k_{{2\mu_2}}\eta_{{\mu_1\mu_3}}{k_2\cdot k_2}}{4\,{ \left(\hat Q\cdot k_2\right)}^{2}}}-{
\frac {{k_1\cdot k_2}\,k_{{1\mu_2}}\eta_{{\mu_1\mu_3}}}{2\, \left(\hat Q\cdot k_1\right) \left(\hat Q\cdot k_2\right)}}+{\frac {5\,{\left(k_1\cdot k_1\right)}^{2}\eta_{{\mu_1\mu_3}}\hat Q_{{\mu_2}}}{8\, \left(\hat Q\cdot k_2\right){ \left(\hat Q\cdot k_1\right)}^{2}}}
$$
$$
-{\frac {k_{{1\mu_2}}\eta_{{\mu_1\mu_3}}{k_1\cdot k_1}}{2\, \left(\hat Q\cdot k_2\right) \left(\hat Q\cdot k_1\right)}}+{\frac {3\,k_{{2\mu_2}}\eta_{{\mu_1\mu_3}}{k_1\cdot k_1}}{16\,{ \left(\hat Q\cdot k_1\right)}^{2}}}+{\frac {\eta_{{\mu_1\mu_3}}\hat Q_{{\mu_2}}{\left(k_1\cdot k_1\right)}^{2}}{16\,{ \left(\hat Q\cdot k_1\right)}^{3}}}
$$
$$
+{\frac {\eta_{{\mu_1\mu_3}}\hat Q_{{\mu_2}}{\left(k_2\cdot k_2\right)}^{2}}{8\, \left(\hat Q\cdot k_1\right){ \left(\hat Q\cdot k_2\right)}^{2}}}+{\frac {k_{{1\mu_2}}\eta_{{\mu_1\mu_3}}{k_2\cdot k_2}}{2\,{ \left(\hat Q\cdot k_2\right)}^{2}}}+{\frac {k_{{1\mu_2}}\eta_{{\mu_1\mu_3}}{k_2\cdot k_2}}{4\, \left(\hat Q\cdot k_2\right) \left(\hat Q\cdot k_1\right)}}
$$
$$
-{\frac {3\,k_{{2\mu_2}}
\eta_{{\mu_1\mu_3}}{k_2\cdot k_2}}{16\, \left(\hat Q\cdot k_2\right) \left(\hat Q\cdot k_1\right)}}+{\frac {3\,k_{{2\mu_2}}\eta_{{\mu_1\mu_3}}{k_1\cdot k_1}}{8\, \left(\hat Q\cdot k_2\right) \left(\hat Q\cdot k_1\right)}}-{\frac {5\,{k_1\cdot k_2}\eta_{{\mu_1\mu_3}}\hat Q_{{\mu_{{2
}}}}{k_2\cdot k_2}}{8\, \left(\hat Q\cdot k_1\right){ \left(\hat Q\cdot k_2\right)}^{2}}}
$$
$$
-{\frac {5\,{k_1\cdot k_1}\eta_{{\mu_1\mu_3}}\hat Q_{{\mu_2}}{k_2\cdot k_2}}{8\, \hat Q\cdot k_
{{1}}{ \left(\hat Q\cdot k_2\right)}^{2}}}+{\frac {5\,{k_1\cdot k_2}\eta_{{\mu_1\mu_3}}\hat Q_{{\mu_2}}{
k_1\cdot k_1}}{8\, \left(\hat Q\cdot k_2\right){ \left(\hat Q\cdot k_1\right)}^{2}}}-{\frac {{k_1\cdot k_1}\eta_{
{\mu_1\mu_3}}\hat Q_{{\mu_2}}{k_2\cdot k_2}}{8\, \left(\hat Q\cdot k_2\right){ \left(\hat Q\cdot k_1\right)
}^{2}}}
$$
$$
+{\frac {k_{{1\mu_3}}\eta_{{\mu_1\mu_2}}{k_1\cdot k_1}}{16\,{ \left(\hat Q\cdot k_1\right)}^{2}}}-{
\frac {3\,k_{{1\mu_3}}k_{{1\mu_1}}\hat Q_{{\mu_2}}{k_1\cdot k_1}}{8\, \left(\hat Q\cdot k_2\right){ 
\left(\hat Q\cdot k_1\right)}^{2}}}+{\frac {9\hat Q_{{\mu_2}}\hat Q_{{\mu_3}}\hat Q_{{\mu_1}}{\left(k_1\cdot k_1\right)}^{3}}{32\,
 \left(\hat Q\cdot k_2\right){ \left(\hat Q\cdot k_1\right)}^{4}}}
$$
$$
+{\frac {9\,k_{{2\mu_3}}\hat Q_{{\mu_2}}\hat Q_{{\mu_1}}{\left(k_2\cdot k_2\right)}^{2}}{32\,{ \left(\hat Q\cdot k_2\right)
}^{3} \left(\hat Q\cdot k_1\right)}}+{\frac {3\,k_{{1\mu_3}}\eta_{{\mu_1\mu_2}}{k_1\cdot k_2}}{8\,
 \left(\hat Q\cdot k_1\right) \left(\hat Q\cdot k_2\right)}}-{\frac {3\,k_{{2\mu_1}}k_{{2\mu_2}}k_{{1\mu
_3}}}{8\, \left(\hat Q\cdot k_1\right) \left(\hat Q\cdot k_2\right)}}
$$
$$
-{\frac {3\,k_{{2\mu_3}}\eta_{{\mu_1\mu_2}}{k_2\cdot k_2}}{16\, \left(\hat Q\cdot k_1\right) \left(\hat Q\cdot k_2\right)}}+{\frac {5\eta_{{\mu_2\mu_3}}{k_2\cdot k_2}\hat Q_{{\mu_1}}{k_1\cdot k_1}}{8\,
 \left(\hat Q\cdot k_2\right){ \left(\hat Q\cdot k_1\right)}^{2}}}+{\frac {\eta_{{\mu_2\mu_3}}
{k_2\cdot k_2}\hat Q_{{\mu_1}}{k_1\cdot k_1}}{8\,{ \left(\hat Q\cdot k_2\right)}^{2} \left(\hat Q\cdot k_1\right)}}
$$
$$
-{\frac {13\,k_{{2\mu_1}}k_{{1\mu_3}}\hat Q_{{\mu_2}}{k_1\cdot k_1}}{16\, \left(\hat Q\cdot k_2\right){
 \left(\hat Q\cdot k_1\right)}^{2}}}-{\frac {5\,{k_1\cdot k_2}\eta_{{\mu_2\mu_3}}\hat Q_{{\mu_1}}
{k_2\cdot k_2}}{8\,{ \left(\hat Q\cdot k_2\right)}^{2} \left(\hat Q\cdot k_1\right)}}-{\frac {3\,k_{{2\mu_3}}k_{{2\,
\mu_2}}\hat Q_{{\mu_1}}{k_1\cdot k_1}}{8\, \left(\hat Q\cdot k_2\right){ \left(\hat Q\cdot k_1\right)}^{2}}}
$$
$$
-{\frac {3\,k_{{1\mu_1}}\eta_{{\mu_2\mu_3}}{k_2\cdot k_2}}{16\,{ \left(\hat Q\cdot k_2\right)}^{2}}}
+{\frac {5\,{k_1\cdot k_2}\eta_{{\mu_2\mu_3}}\hat Q_{{\mu_1}}{k_1\cdot k_1}}{8\, \hat Q\cdot k_
{{2}}{ \left(\hat Q\cdot k_1\right)}^{2}}}+{\frac {13\,k_{{2\mu_3}}k_{{1\mu_2}}\hat Q_{{\mu_1}}
{k_2\cdot k_2}}{16\,{ \left(\hat Q\cdot k_2\right)}^{2} \left(\hat Q\cdot k_1\right)}}
$$
$$
+{\frac {9\,k_{{1\mu_3}}\hat Q_{{\mu_2}}\hat Q_{{\mu_1}}{k_2\cdot k_2}\,{k_1\cdot k_1}}{32\,{ \left(\hat Q\cdot k_2\right)}^{2}{ \left(\hat Q\cdot k_1\right)}^{2}}}-{\frac {\eta_{{\mu_1\mu_2}}{k_2\cdot k_2}\hat Q_{
{\mu_3}}{k_1\cdot k_1}}{8\, \left(\hat Q\cdot k_2\right){ \left(\hat Q\cdot k_1\right)}^{2}}}-{\frac {3\,k_{{1\,
\mu_1}}k_{{1\mu_2}}\hat Q_{{\mu_3}}{k_2\cdot k_2}}{16\,{ \left(\hat Q\cdot k_2\right)}^{2} \left(\hat Q\cdot k_1\right)}}
$$
$$
+{\frac {3\,k_{{1\mu_3}}k_{{1\mu_1}}\hat Q_{{\mu_2}}{k_2\cdot k_2}}{8\,{ \left(\hat Q\cdot k_2\right)}^{2}
 \left(\hat Q\cdot k_1\right)}}+{\frac {\eta_{{\mu_2\mu_3}}k_{{2\mu_1}}{k_2\cdot k_2}}{2\, \left(\hat Q\cdot k_1\right) \left(\hat Q\cdot k_2\right)}}+{\frac {3\,k_{{2\mu_3}}\eta_{{\mu_1\mu_2}}
{k_1\cdot k_1}}{16\,{ \left(\hat Q\cdot k_1\right)}^{2}}}
$$
$$
-{\frac {k_{{1\mu_1}}\eta_{{\mu_2\mu_3}}{k_1\cdot k_1}}{4\,{ \left(\hat Q\cdot k_1\right)}^{2}}}-{
\frac {3\,k_{{2\mu_1}}k_{{2\mu_2}}\hat Q_{{\mu_3}}{k_2\cdot k_2}}{16\,{ \left(\hat Q\cdot k_2\right)}^{2}
 \left(\hat Q\cdot k_1\right)}}+{\frac {9\,k_{{2\mu_2}}\hat Q_{{\mu_1}}\hat Q_{{\mu_3}}{\left(k_1\cdot k_1\right)}^{2}}{32
\, \left(\hat Q\cdot k_2\right){ \left(\hat Q\cdot k_1\right)}^{3}}}
$$
$$
+{\frac {13\,k_{{2\mu_1}}k_{{1\mu_3}}\hat Q_{{\mu_2}}{k_2\cdot k_2}}{16\,{ \left(\hat Q\cdot k_2\right)}^{
2} \left(\hat Q\cdot k_1\right)}}+{\frac {\eta_{{\mu_1\mu_2}}\hat Q_{{\mu_3}}{\left(k_1\cdot k_1\right)}^{2}}{16\,{
 \left(\hat Q\cdot k_1\right)}^{3}}}-{\frac {\eta_{{\mu_2\mu_3}}{\left(k_1\cdot k_1\right)}^{2}\hat Q_{{\mu_1}}}{8\,
 \left(\hat Q\cdot k_2\right){ \left(\hat Q\cdot k_1\right)}^{2}}}
$$
$$
+{\frac {k_{{1\mu_3}}k_{{2\mu_2}}\hat Q_{{\mu_1}}{k_2\cdot k_2}}{8\,{ \left(\hat Q\cdot k_2\right)}^{2}
 \left(\hat Q\cdot k_1\right)}}-{\frac {7\,{k_1\cdot k_2}\eta_{{\mu_1\mu_2}}\hat Q_{{\mu_3}}
{k_1\cdot k_1}}{8\, \left(\hat Q\cdot k_2\right){ \left(\hat Q\cdot k_1\right)}^{2}}}+{\frac {7\,{k_1\cdot k_2}\eta_{{
\mu_1\mu_2}}\hat Q_{{\mu_3}}{k_2\cdot k_2}}{8\,{ \left(\hat Q\cdot k_2\right)}^{2} \hat Q\cdot k_{
{1}}}}
$$
$$
-{\frac {9\,k_{{2\mu_2}}\hat Q_{{\mu_1}}\hat Q_{{\mu_3}}{k_2\cdot k_2}\,{k_1\cdot k_1}}{32\,{ \left(\hat Q\cdot k_2\right)}^{2}{ \left(\hat Q\cdot k_1\right)}^{2}}}+{\frac {9\,k_{{2\mu_2}}\hat Q_{{\mu_1}}\hat Q_{{\mu_{{3}
}}}{\left(k_2\cdot k_2\right)}^{2}}{32\,{ \left(\hat Q\cdot k_2\right)}^{3} \left(\hat Q\cdot k_1\right)}}-{\frac {9\,k_{{1\,
\mu_3}}\hat Q_{{\mu_2}}\hat Q_{{\mu_1}}{\left(k_2\cdot k_2\right)}^{2}}{32\,{ \left(\hat Q\cdot k_2\right)}^{3} \left(\hat Q\cdot k_1\right)}}
$$
$$
+{\frac {k_{{2\mu_1}}\eta_{{\mu_2\mu_3}}{k_1\cdot k_2}}{2\, \left(\hat Q\cdot k_1\right) \left(\hat Q\cdot k_2\right)}}-{\frac {k_{{2\mu_1}}k_{{1\mu_2}}\hat Q_{{\mu_3}}{k_2\cdot k_2}}
{{ \left(\hat Q\cdot k_2\right)}^{2} \left(\hat Q\cdot k_1\right)}}+{\frac {3\,k_{{1\mu_3}}\eta_{{\mu_1\mu_2}}{k_1\cdot k_1
}}{16\, \left(\hat Q\cdot k_1\right) \left(\hat Q\cdot k_2\right)}}
$$
$$
-{\frac {3\,k_{{1\mu_3}}\eta_{{\mu_1\mu_2}}{k_2\cdot k_2}}{16\,{ \left(\hat Q\cdot k_2\right)}^{2}}}
-{\frac {\eta_{{\mu_2\mu_3}}k_{{2\mu_1}}{k_1\cdot k_1}}{2\,{ \left(\hat Q\cdot k_1\right)}^{2}}}+{
\frac {\eta_{{\mu_1\mu_2}}{\left(k_2\cdot k_2\right)}^{2}\hat Q_{{\mu_3}}}{8\,{ \left(\hat Q\cdot k_2\right)}^{2}
 \left(\hat Q\cdot k_1\right)}}
$$
$$
+{\frac {3\,k_{{2\mu_3}}k_{{2\mu_2}}k_{{1\mu_1}}}{8\, \left(\hat Q\cdot k_1\right) 
\left(\hat Q\cdot k_2\right)}}-{\frac {3\,k_{{1\mu_1}}\eta_{{\mu_2\mu_3}}{k_2\cdot k_2}}{8\, \left(\hat Q\cdot k_1\right) \left(\hat Q\cdot k_2\right)}}-{\frac {3\,k_{{2\mu_3}}{k_1\cdot k_2}\eta_{{\mu_1\mu_{
{2}}}}}{8\, \left(\hat Q\cdot k_1\right) \left(\hat Q\cdot k_2\right)}}
$$
$$
+{\frac {3\,k_{{2\mu_3}}\eta_{{\mu_1\mu_2}}{k_1\cdot k_1}}{16\, 
\left(\hat Q\cdot k_1\right) \left(\hat Q\cdot k_2\right)}}+{\frac {3\,k_{{2\mu_3}}k_{{1\mu_2}}k_{{1\mu_1}}}{8\, \hat Q\cdot k_
{{1}} \left(\hat Q\cdot k_2\right)}}-{\frac {\eta_{{\mu_2\mu_3}}k_{{2\mu_1}}{k_1\cdot k_1}}{4\,
 \left(\hat Q\cdot k_1\right) \left(\hat Q\cdot k_2\right)}}
$$
$$
-{\frac {3\,k_{{1\mu_3}}\eta_{{\mu_1\mu_2}}{k_2\cdot k_2}}{16\, 
\left(\hat Q\cdot k_1\right) \left(\hat Q\cdot k_2\right)}}-{\frac {3\,k_{{1\mu_3}}k_{{2\mu_2}}k_{{1\mu_1}}}{8\, \hat Q\cdot k_
{{1}} \left(\hat Q\cdot k_2\right)}}-{\frac {k_{{2\mu_3}}k_{{1\mu_1}}\hat Q_{{\mu_2}}{k_1\cdot k_1}}{8\,
 \left(\hat Q\cdot k_2\right){ \left(\hat Q\cdot k_1\right)}^{2}}}
$$
$$
+{\frac {k_{{2\mu_3}}k_{{1\mu_1}}\hat Q_{{\mu_2}}{k_2\cdot k_2}}{8\,{ \left(\hat Q\cdot k_2\right)}^{2}
 \left(\hat Q\cdot k_1\right)}}+{\frac {3\,k_{{1\mu_1}}k_{{1\mu_2}}\hat Q_{{\mu_3}}{k_1\cdot k_1}}{16\,
 \left(\hat Q\cdot k_2\right){ \left(\hat Q\cdot k_1\right)}^{2}}}+{\frac {3\,k_{{2\mu_1}}k_{{2\mu_2}}\hat Q_{
{\mu_3}}{k_1\cdot k_1}}{16\, \left(\hat Q\cdot k_2\right){ \left(\hat Q\cdot k_1\right)}^{2}}}
$$
$$
-{\frac {\eta_{{\mu_1\mu_2}}{\left(k_1\cdot k_1\right)}^{2}\hat Q_{{\mu_3}}}{8\, 
\left(\hat Q\cdot k_2\right){ \left(\hat Q\cdot k_1\right)}^{2}}}+{\frac {\eta_{{\mu_1\mu_2}}{k_1\cdot k_1}\hat Q_{{\mu_3}}{k_2\cdot k_2}}{8
\,{ \left(\hat Q\cdot k_2\right)}^{2} \left(\hat Q\cdot k_1\right)}}+{\frac {k_{{2\mu_1}}k_{{1\mu_2}}\hat Q_{{
\mu_3}}{k_1\cdot k_1}}{ \left(\hat Q\cdot k_2\right){ \left(\hat Q\cdot k_1\right)}^{2}}}
$$
$$
-{\frac {5\eta_{{\mu_2\mu_3}}{\left(k_2\cdot k_2\right)}^{2}\hat Q_{{\mu_1}}}{8\,{ \left(\hat Q\cdot k_2\right)}^{2
} \left(\hat Q\cdot k_1\right)}}-{\frac {13\,k_{{2\mu_3}}k_{{1\mu_2}}\hat Q_{{\mu_1}}{k_1\cdot k_1}}{16\,
 \left(\hat Q\cdot k_2\right){ \left(\hat Q\cdot k_1\right)}^{2}}}-{\frac {k_{{1\mu_3}}k_{{2\mu_2}}\hat Q_{{
\mu_1}}{k_1\cdot k_1}}{8\, \left(\hat Q\cdot k_2\right){ \left(\hat Q\cdot k_1\right)}^{2}}}
$$
$$
-{\frac {k_{{2\mu_3}}\eta_{{\mu_1\mu_2}}{k_2\cdot k_2}}{16\,{ \left(\hat Q\cdot k_2\right)}^{2}}}+{
\frac {9\,k_{{2\mu_3}}\hat Q_{{\mu_2}}\hat Q_{{\mu_1}}{\left(k_1\cdot k_1\right)}^{2}}{32\, \left(\hat Q\cdot k_2\right){
 \left(\hat Q\cdot k_1\right)}^{3}}}-{\frac {9\,k_{{2\mu_3}}\hat Q_{{\mu_2}}\hat Q_{{\mu_1}}{k_2\cdot k_2}\,{
k_1\cdot k_1}}{32\,{ \left(\hat Q\cdot k_2\right)}^{2}{ \left(\hat Q\cdot k_1\right)}^{2}}}
$$
$$
-{\frac {9\hat Q_{{\mu_2}}\hat Q_{{\mu_3}}\hat Q_{{\mu_1}}{k_2\cdot k_2}\,{\left(k_1\cdot k_1\right)}^{2}}{32\,{ \left(\hat Q\cdot k_2\right)}^{2}{ \left(\hat Q\cdot k_1\right)}^{3}}}+{\frac {9\hat Q_{{\mu_2}}\hat Q_{{\mu_3}}\hat Q_{{\mu_1}
}{\left(k_2\cdot k_2\right)}^{2}{k_1\cdot k_1}}{32\,{ \left(\hat Q\cdot k_2\right)}^{3}{ \left(\hat Q\cdot k_1\right)}^{2}}}-{
\frac {9\hat Q_{{\mu_2}}\hat Q_{{\mu_3}}\hat Q_{{\mu_1}}{\left(k_2\cdot k_2\right)}^{3}}{32\,{ \left(\hat Q\cdot k_2\right)}^{
4} \left(\hat Q\cdot k_1\right)}}
$$
$$
+{\frac {3\,k_{{1\mu_3}}k_{{1\mu_2}}\hat Q_{{\mu_1}}{k_1\cdot k_1}}{16\, \left(\hat Q\cdot k_2\right){
 \left(\hat Q\cdot k_1\right)}^{2}}}-{\frac {3\,k_{{1\mu_3}}k_{{1\mu_2}}\hat Q_{{\mu_1}}{k_2\cdot k_2}}{
16\,{ \left(\hat Q\cdot k_2\right)}^{2} \left(\hat Q\cdot k_1\right)}}-{\frac {\eta_{{\mu_1\mu_2}}\hat Q_{{\mu_
{{3}}}}{\left(k_2\cdot k_2\right)}^{2}}{16\,{ \left(\hat Q\cdot k_2\right)}^{3}}}
$$
$$
-{\frac {9\,k_{{1\mu_1}}\hat Q_{{\mu_2}}\hat Q_{{\mu_3}}{\left(k_1\cdot k_1\right)}^{2}}{32\, \left(\hat Q\cdot k_2\right)
{ \left(\hat Q\cdot k_1\right)}^{3}}}+{\frac {9\,k_{{1\mu_1}}\hat Q_{{\mu_2}}\hat Q_{{\mu_3}}{k_2\cdot k_2}\,{
k_1\cdot k_1}}{32\,{ \left(\hat Q\cdot k_2\right)}^{2}{ \left(\hat Q\cdot k_1\right)}^{2}}}-{\frac {9\,k_{{1\mu_{{
1}}}}\hat Q_{{\mu_2}}\hat Q_{{\mu_3}}{\left(k_2\cdot k_2\right)}^{2}}{32\,{ \left(\hat Q\cdot k_2\right)}^{3} 
\left(\hat Q\cdot k_1\right)}}
$$
$$
-{\frac {k_{{2\mu_2}}k_{{1\mu_1}}\hat Q_{{\mu_3}}{k_1\cdot k_1}}{4\, 
\left(\hat Q\cdot k_2\right){ \left(\hat Q\cdot k_1\right)}^{2}}}+{\frac {k_{{2\mu_2}}k_{{1\mu_1}}\hat Q_{{\mu_3}}{k_2\cdot k_2}}{4\,{ \left(\hat Q\cdot k_2\right)}^{2} \left(\hat Q\cdot k_1\right)}}-{\frac {9\,k_{{1\mu_3}}\hat Q_{{\mu_2}}\hat Q_{{\mu_1
}}{\left(k_1\cdot k_1\right)}^{2}}{32\, \left(\hat Q\cdot k_2\right){ \left(\hat Q\cdot k_1\right)}^{3}}}
$$
$$
-{\frac {\eta_{{\mu_2\mu_3}}\hat Q_{{\mu_1}}{\left(k_2\cdot k_2\right)}^{2}}{16\,{ \left(\hat Q\cdot k_2\right)}^{3}}
}+{\frac {3\,k_{{1\mu_1}}\eta_{{\mu_2\mu_3}}{k_1\cdot k_1}}{16\, 
\left(\hat Q\cdot k_2\right) \left(\hat Q\cdot k_1\right)}}+{\frac {3\,k_{{2\mu_3}}k_{{2\mu_2}}\hat Q_{{\mu_1}}{k_2\cdot k_2}}{8\,{ \left(\hat Q\cdot k_2\right)}^{2} \left(\hat Q\cdot k_1\right)}}
$$
$$
+{\frac {3\,k_{{2\mu_{{3}}}}k_{{2\mu_{{1}}}}\hat Q_{{\mu_{{2}}}}{k_1\cdot k_1}}{16\, 
\left(\hat Q\cdot k_2\right){
\left(\hat Q\cdot k_1\right)}^{2}}}
-{\frac {3\,k_{{2\mu_{{3}}}}k_{{2\mu_{{1}}}}\hat Q_{{\mu_{{2}}}}{k_2\cdot k_2}}{
16\,{ \left(\hat Q\cdot k_2\right)}^{2} 
\left(\hat Q\cdot k_1\right)}}
$$
$$
+ \left\{ \; \mbox{cyclic permutations of} \; (k_1,\mu_1),
(k_2,\mu_2), (k_3,\mu_3)\; \right\}
$$




\bigskip
\bigskip
\bigskip
\begin{thebibliography}{10}

\centerline{\bf REFERENCES}

\bibitem{Weldon}
A. H. Weldon, Phys. Rev. D {\bf 26}, 1394 (1982).

\bibitem{KajantieKapusta}
K. Kajantie and J. Kapusta, Ann Phys, {\bf 160}, 477 (1985).

\bibitem{Pisarski}
R. D. Pisarski, Nucl. Phys. {\bf B309}, 476 (1988).

\bibitem{BraatenPisarski92}
E. Braaten and R. D. Pisarski, Phys. Rev. D {\bf 45}, 1827 (1992).

\bibitem{FrenkelTaylor90} 
J. Frenkel and J. C. Taylor, Nucl. Phys. {\bf B334}, 199 (1990);
{\bf B374}, 156 (1992).
\bibitem{BraatenPisarski90}
E. Braaten and R. D. Pisarski, Nucl. Phys. {\bf B337}, 569 (1990);
{\bf B339}, 310 (1990).
\bibitem{Kapusta}
J. Kapusta, {\it Finite Temperature Field Theory}
Cambridge University Press, Cambridge, England, 1989)

\bibitem{BrandtFrenkelTaylor91}
F. T. Brandt, J. Frenkel and J. C. Taylor,
Phys. Rev. D {\bf 44}, 1801 (1991).
\bibitem{NakkagawaNiegawa} 
H. Nakkagawa and A. Niegawa,
Phys. Lett. B {\bf 193}, 263 (1987).

\bibitem{FujimotoYamada}
Y. Fujimoto and H. Yamada,
Phys. Lett. B {\bf 200}, 167 (1988).

\bibitem{BaierPireSchiff}
R. Baier, B. Pire and D. Schiff,
Z. Phys. C {\bf 51}, 581 (1991).

\bibitem{EijckStephensWeert}
M. A. van Eijck, C. R. Stephens and C. van der Weert,
Mod. Phys. Lett. A9, 309 (1994).

\bibitem{Barton}
G. Barton, Ann. Phys. 200, 271 (1990).

\bibitem{DavydychevOsland}
A.I. Davydychev, P. Osland and O.V. Tarasov,
Phys. Rev. D {\bf 54}, 4087 (1996).

\bibitem{BallChiu} 
S. Ball and  Ting-Wai Chiu,
Phys. Rev. D {\bf 22}, 2550 (1981).

\bibitem{BrandtFrenkel}
F. T. Brandt and J. Frenkel,
Phys. Rev. Lett. {\bf 74}, 1705 (1995).

\bibitem{Gradshteyn}  I. S. Gradshteyn and I. Ryzhik, 
{\it Table of Integrals, Series and Products} 
(Academic Press, New York, 1980).

\bibitem{BrandtFrenkelTaylor94}
F. T. Brandt, J. Frenkel and J. C. Taylor,
Phys. Rev. D {\bf 50}, 411 (1994).

\bibitem{PerryCollins}
J. Perry and J. C. Collins,
Phys. Rev. Lett. {\bf 34}, 1353 (1975).
\bibitem{MatsumotoNakanoUmezawa}
H. Matsumoto, Y. Nakano and H. Umezawa,
Phys. Rev. D {\bf 29}, 1116 (1984).
\bibitem{ElmforsKobes}
P. Elmfors and R. Kobes,
Phys Rev. D {\bf 51}, 774 (1995).
\bibitem{ChaichianHayashi}
M. Chaichian and M. Hayashi,
Acta Phys. Polon. {\bf 27}, (1996). 

\bibitem{FrenkelSaaTaylor}
J. Frenkel, A. Saa and J. C. Taylor,
Phys. Rev. D {\bf 46}, 3670 (1992).
\bibitem{ArnoldZhai}
P. Arnold and  C. Zhai,
Phys. Rev. D {\bf 51}, 1906 (1995).
\bibitem{BraatenNieto}
E. Braaten and A. Nieto,
Phys. Rev. D {\bf 53}, 1996 (1996). 

\end{thebibliography}
\end{document}